\documentclass[12pt,a4paper]{article}\usepackage[]{graphicx}\usepackage[]{color}
\makeatletter
\def\maxwidth{ %
  \ifdim\Gin@nat@width>\linewidth
    \linewidth
  \else
    \Gin@nat@width
  \fi
}
\makeatother

\definecolor{fgcolor}{rgb}{0.345, 0.345, 0.345}
\newcommand{\hlstr}[1]{\textcolor[rgb]{0.192,0.494,0.8}{#1}}%
\newcommand{\hlopt}[1]{\textcolor[rgb]{0,0,0}{#1}}%
\newcommand{\hlstd}[1]{\textcolor[rgb]{0.345,0.345,0.345}{#1}}%
\newcommand{\hlkwd}[1]{\textcolor[rgb]{0.737,0.353,0.396}{\textbf{#1}}}%

\usepackage{framed}
\makeatletter
\newenvironment{kframe}{%
 \def\at@end@of@kframe{}%
 \ifinner\ifhmode%
  \def\at@end@of@kframe{\end{minipage}}%
  \begin{minipage}{\columnwidth}%
 \fi\fi%
 \def\FrameCommand##1{\hskip\@totalleftmargin \hskip-\fboxsep
 \colorbox{shadecolor}{##1}\hskip-\fboxsep
     \hskip-\linewidth \hskip-\@totalleftmargin \hskip\columnwidth}%
 \MakeFramed {\advance\hsize-\width
   \@totalleftmargin\z@ \linewidth\hsize
   \@setminipage}}%
 {\par\unskip\endMakeFramed%
 \at@end@of@kframe}
\makeatother

\definecolor{shadecolor}{rgb}{.97, .97, .97}
\definecolor{messagecolor}{rgb}{0, 0, 0}
\definecolor{warningcolor}{rgb}{1, 0, 1}
\definecolor{errorcolor}{rgb}{1, 0, 0}
\newenvironment{knitrout}{}{} 

\usepackage{alltt}

\usepackage{amssymb}
\usepackage{amsmath}
\usepackage{algpseudocode}
\usepackage{algorithm}
\usepackage{algpseudocode}

\usepackage{float}
\usepackage[english]{babel}
\usepackage{rotating}
\usepackage{caption}
\numberwithin{equation}{section}

\usepackage{etoolbox} \makeatletter \preto{\@verbatim}{\topsep=0pt \partopsep=0pt } \makeatother

\usepackage{epsf}
\usepackage{color}
\usepackage{tikz}

\usepackage{hyperref}

\usepackage{natbib}
\bibpunct[, ]{(}{)}{;}{a}{,}{,}

\IfFileExists{upquote.sty}{\usepackage{upquote}}{}
\begin{document}

\title{Assurance for clinical trial design with normally distributed outcomes: eliciting uncertainty about variances}
\author{Ziyad A. Alhussain\textsuperscript{1}, Jeremy E. Oakley\textsuperscript{2}}
\maketitle
\begin{center}
 \vspace{-.5em}
 {\small \textsuperscript{1} Mathematics Department, Faculty of Science in Zulfi, Majmaah University, Saudi Arabia  \\[-.3em]
 {\tt z.alhussain@mu.edu.sa}}
 \vspace{.1em}

{\small \textsuperscript{2} School of Mathematics and Statistics, University of Sheffield, UK\\[-.3em]
{\tt j.oakley@sheffield.ac.uk}}

\end{center}


\begin{abstract}
The assurance method is growing in popularity in clinical trial planning. The method involves eliciting a prior distribution for the treatment effect, and then calculating the probability that a proposed trial will produce a `successful' outcome. For normally distributed observations, uncertainty about the variance of the normal distribution also needs to be accounted for, but there is little guidance in the literature on how to elicit a distribution for a variance parameter. We present a simple elicitation method, and illustrate how the elicited distribution is incorporated within an assurance calculation. We also consider multi-stage trials, where a decision to proceed with a larger trial will follow from the outcome of a smaller trial; we illustrate the role of the elicted distribution in assessing the information provided by a proposed smaller trial. Free software is available for implementing our methods. 
\end{abstract}


\section{Introduction}
Assurance is a Bayesian alternative to a power calculation for choosing a sample size in a clinical trial. A prior distribution is elicited for the treatment effect, and  the prior probability that the trial will be `successfull' is calculated, for any success criteria that the trial sponsor wishes to consider (e.g. that the observed treatment effect will be positive, and statistically significant at the appropriate size). This approach was first proposed by \cite{SpiegelhalterFreedman:1986}, and developed in \cite{O'Hagan:2001} and \cite{O'Hagan:2005}, where the term ``assurance" was used. There is growing interest in the assurance method, with its use on a large scale described in \cite{Dallow:2018} and \cite{Crisp:2018}.

The aim of the assurance method is to provide a realistic assessment of the trial sponsor's probability of a successful trial. This requires a prior distribution for any relevant uncertain quantity. \cite{O'Hagan:2005} consider the case of prior distributions for normally distributed and binomial data, and \cite{Ren:2014} presented elicitation methodology for time-to-event data. \cite{Gasparini:2013} also considered normally distributed data. Neither O'Hagan et al. nor Gasparini et al. discuss how one would elicit distributions for unknown population variances, and there is little guidance in the wider elicitation literature on how to elicit a distribution for a variance. We propose a method in this paper, and show how to incorporate it within an assurance calculation.

In cases where there is greater prior uncertainty, trial sponsors may consider an adaptive design, or a multi-stage approach, where planning decisions about a latter stage trial may be informed by the results from an earlier stage. A case-study involving the use of assurance in such a setting is given in \cite{Nixon:2009}. For normally distributed data, it is again important to consider uncertainty in the variance, as the variance will affect how much information one gains for a given sample size. We illustrate how simulation can be used to investigate how a small study would reduce uncertainty about a treatment effect, to support the planning of a larger trial.

In the next section, we discuss the assurance method and note the role of variance distributions in assurance calculations. In Section 3, we review methods for eliciting a distribution for the treatment effect, and discuss elicitation of distributions for variances in Section 4. In Section 5, we show how the elicited distributions are incorporated within the assurance calculation, and we discuss the extension to multi-stage trial planning in Section 6. Free software is available to implement all our methods, and is described in the Appendix.

\section{Assurance}

In this section, we describe the assurance method, specifically in the context of a randomised controlled trial, where the observations in both the treatment and control arms are assumed to be normally distributed. We suppose that in the control arm, we have observations $X_1,\ldots,X_{n_c}\stackrel{i.i.d.}{\sim}N(\mu_c, \sigma^2_c)$, and in the treatment arm, we have observations  $Y_1,\ldots,Y_{n_t}\stackrel{i.i.d.}{\sim}N(\mu_t, \sigma^2_t)$. We write $\mu_t = \mu_c + \delta$, so that we interpret $\delta$ as the treatment effect. 

\cite{O'Hagan:2005} consider the four cases of a one-sided superiority trial, a two-sided superiority trial, a non-inferiority trial and an equivalence trial. In this paper we will consider two-sided superiority trials only, but extension to the other cases would be straightforward (and would not change the methodology we are proposing here).

The data will be analysed with a two-sample $t$-test:
$$
T = \frac{\bar{Y} - \bar{X}}{\sqrt{\frac{S^2_X}{n_c} + \frac{S^2_Y}{n_t}}},
$$
with $T$ compared with the Student-$t$ distribution with $\nu$ degrees of freedom computed using the Welch approximation. 

For a power calculation, for a given $n_t$ and $n_c$, we would fix values of $\delta, \sigma^2_t$ and $\sigma^2_c$, and calculate the probability of observing data such that the null hypothesis is rejected, for a specified level of significance. We denote $R$ to be the event of rejecting the null hypothesis, and write this probability as
$$
Pr(R|\delta, \sigma^2_t,\sigma^2_c).
$$
(In practice, we might make some simplifications such as assuming $\sigma^2_t=\sigma^2_c$, and that $T$ can be compared with the standard normal distribution). In this calculation, we interpret $\delta$ as a `minimum clinically relevant' treatment effect: the smallest treatment effect that we would want our trial to be able to detect.

In the assurance method, we consider the unconditional probability of the \textit{same} event $R$, but with a \textit{different} interpretation of $\delta$: we now interpret $\delta$ as the \textit{true} treatment effect, elicit a  prior distribution $\pi(\delta, \sigma^2_t, \sigma^2_c)$, and compute
$$
Pr(R) = \int Pr(R|\delta, \sigma^2_t,\sigma^2_c)\pi(\delta, \sigma^2_t, \sigma^2_c)d\delta d\sigma^2_t d\sigma^2_c.
$$
We emphasise that the event $R$ is the same as that used in the power calculation: although we now have a prior distribution $\pi(\delta, \sigma^2_t, \sigma^2_c)$ we do \textit{not} assume it will be used in the analysis of the trial data: we assume exactly the same (frequentist) analysis as that used in the power calculation. In general, we can think of a regulator or trial sponsor specifying an event $R$ in which the trial outcome is ``successful'', and then we elicit a prior distribution \textit{only} to assess the probability of achieving the ``successful'' outcome.

The computation of $Pr(R)$ is usually straightforward using Monte Carlo methods; the main effort required is in eliciting the prior $\pi(\delta, \sigma^2_t, \sigma^2_c)$ which we will discuss in the following sections.

\section{Eliciting a prior distribution for the treatment effect\label{sec:uniElicitation}}

For simplicity and ease of exposition, it is supposed that there is one female expert, and that the elicitation is conducted by a male facilitator. There are various general considerations when performing elicitation such as training of the experts, and how to manage (or combine opinions from) multiple experts. The focus of this paper is solely on how to elicit judgements about a mean and variance, and we do not consider these other issues here. Guidance on these and other aspects of elicitation can be found in \cite{Dias:2018}, \cite{EFSA:2014}, \cite{Hagan:2006}, \cite{Cooke:1991} and \cite{Morgan:1990}.

We suppose that an expert would be equally willing to consider the treatment effect $\delta$ directly, or to consider the mean in the treatment group $\mu_t$, given a hypothetical value for the mean in the control group $\mu_c$ (the expert might propose her own hypothetical value for $\mu_c$, or an estimate may be available from previous trials with the same control arm). In the following discussion, we consider the former option. 

General advice in elicitation methods is to ask experts about observable quantities, rather than parameters in statistical models \cite[see, for example,][]{Kadane:1998}. Although not strictly observable, we think the mean of a normal distribution would be well-enough understood for an expert to make judgements about it directly. Hence, standard univariate elicitation methods can be used to elicit a prior distribution for $\delta$ or $\mu_t|\mu_c$ \citep[see, for example, ][]{Oakley:2010}.

Such methods typically involve eliciting a small number of points from the expert's cumulative distribution function of $\delta$: the expert judges $Pr(\delta \le d_i) = p_i$, for $i = 1, \ldots, n$. We can specify $d_1,\ldots, d_n$ and ask the expert to provide $p_1,\ldots, p_n$, or vice-versa. For example, the expert can be asked to provide her quartiles, in which case $p_1, p_2, p_3$ are fixed at 0.25, 0.5 and 0.75 respectively, and the expert provides the corresponding values of $d_1,d_2,d_3$.

We then consider some parametric family of distributions, indexed by parameters $\theta$, and choose $\theta$ to minimise 
$$
\sum_{i=1}^n \left(F(d_i;\theta) - p_i \right)^2,
$$
where $F(.;\theta)$ is the cumulative distribution function from the chosen family with parameter $\theta$. Both \cite{Gasparini:2013} and \cite{O'Hagan:2005} assume a normal distribution $\delta\sim N(m,v)$, so that we would have $\theta=(m, v)$. This approach can be implemented in R \citep{R2019} using the package \texttt{SHELF} \citep{OakleySHELF}, and is incorporated in our software.

\subsection{Mixture distributions}

\cite{O'Hagan:2005} also consider a mixture distribution where a non-zero probability is given to the event $\delta = 0$: the event that the treatment has no effect, and a conditional distribution is elicited for $\delta$ given that $\delta\neq 0$: the treatment has some effect. This 
approach is also used in \cite{Dallow:2018}, and we have implemented this in our software.

To see the possible benefits of the mixture approach, consider the following example. Suppose an expert judges a 70\% chance that the new treatment will have some (beneficial) effect, and also thinks that a treatment effect of 0.5 (on some appropriate scale) is `most likely'. If we suppose $\delta \sim N(0.5, 1)$, then we have a mode at 0.5 and $Pr(\delta>0) \simeq 0.7$: the $N(0.5, 1)$ distribution would appear to describe well the two judgements made by the expert. However, this distribution would also imply $Pr(\delta >1)\simeq 0.3$, but the expert may not judge a treatment effect twice as high as her `most likely value' to be very plausible, and giving such weight to larger values of $\delta$ may adversely effect the assessment of the required sample size. Using the mixture approach, we could instead set $Pr(\delta = 0)=0.3$, and then, conditional on $\delta \neq 0$, set $\delta \sim N(0.5, v)$, with a smaller, more appropriate choice of $v$.

\section{Eliciting a distribution for a variance}

To the best of our knowledge, there has been little work on eliciting beliefs about variances. One existing approach that can be used is based on eliciting beliefs about parameters in linear regression models. \cite{KADANE:1980} and \cite{Al-Awadhi:1998} consider elicitation for  the parameters $(\mu,\beta_1,\ldots,\beta_p,\sigma^2)$ in regression models of the form
\begin{equation*}
  X_i = \mu + \sum_{j=1}^{p} \beta_{j} z_{ij} + \varepsilon_i, \hspace{0.2 in} \mbox{for} \hspace{ 0.1 in} i = 1,2,...,n,
\end{equation*}
where $\varepsilon_1, ..., \varepsilon_n \stackrel{i.i.d}{\sim}  \mbox{N} (0, \sigma^2)$. By setting $\beta_j=0$ for all $j$, this would reduce to our case. \cite{Al-Awadhi:1998} proposed an elicitation method for quantifying opinions about the parameters of a multivariate normal distribution; the same elicitation method could be used for quantifying beliefs about a univariate normally distributed population. These methods require the expert to update her judgements in light of hypothetical data, under the assumption that the expert updates  her beliefs using Bayes' theorem. We think this is a difficult task: the expert may not view hypothetical data as credible and behave the same way had she observed real data, and it is unlikely that the expert would weight prior knowledge and hypothetical data precisely according to Bayes' theorem in any case. The expert may be insensitive to the sample size, for example in accounting for the variability in a sample mean \cite[]{Tversky:1971}. We think it desirable to have alternative elicitation methods available to the expert.

 \cite{KADANE:1980} and \cite{Al-Awadhi:1998} infer judgements about the parameters $\mu$ and $\sigma^2$ from judgements about the observable quantities $X_i$, by eliciting summaries from the expert's predictive distribution. For example, suppose we wish to elicit an expert's opinion about the variance parameter $\sigma^2$ of a random variable $X$ that follows a normal distribution with a known mean $\mu$. Since $\sigma^2$ is not directly observable then the expert is asked to make judgements about the observable quantity $X$, and we infer $p(\sigma^2)$ from these judgements. Any choice of $p(\sigma^2)$ implies a distribution
\begin{equation*}
    p_X(x) = \int_{\mathbb{R}^{+}} p_X(x \mid \sigma^2) p(\sigma^2) d\sigma^2,
\end{equation*}
and we suppose that a particular choice of $p(\sigma^2)$ will result in the above integral (approximately) matching the expert's beliefs about $X$, so that this choice of $p(\sigma^2)$ describes the expert's underlying beliefs about $\sigma^2$. A concern here is whether an expert really is able to account for her uncertainty about $\sigma^2$ when making judgements about $X$. A possibility is that the expert instead only makes judgements about $X$ conditional on some estimate of $\sigma^2$.

\cite{KADANE:1980} use conjugate priors for $\mu$ and $\sigma^2$ which force the expert's opinion about the two parameters to be dependent. However, it is possible in reality that knowledge of one parameter would not change the expert's opinion about the other. \cite{Al-Awadhi:2001}  argued that, unless mathematical tractability is required, then it can be better to assume independence between the two parameters, and that this helps the expert focus on the assessments of each parameter separately. They proposed an elicitation method for the multivariate normal distribution where the mean vector and covariance matrix are assumed to be independent, though their method also asks the expert to update her judgements in the light of hypothetical data.

We argue that the better informed the expert, the less likely a judgement of dependence between the two parameters would be required. For example, consider the distribution of running times for an individual over a distance of 5km. With no information about the ability of the runner, one might have considerable uncertainty about the mean, e.g.\ an interval of 15 minutes to 1 hour may be judged plausible, with smaller variances of running times associated with smaller means within this interval. But if one already has good knowledge about the particular runner's ability, a much smaller interval may be judged plausible for the mean, and one's beliefs about the variance may not change appreciably given different plausible means.

We propose a new elicitation method for quantifying opinions about an uncertain population mean and variance. Our  method does not elicit judgements using hypothetical data and Bayes' theorem, it does not use predictive elicitation and it assumes independence between the mean and variance.

\subsection{The proposed elicitation method}\label{sec:method}

We first ask the expert to suppose that the treatment works and that the effect is `as expected': more precisely, we suppose that $\delta=m$, where $m$ could be the mean or median of the expert's distribution for $\delta$ (where, in the mixture case, the mean/median would be conditional on $\delta \neq 0$). The expert can choose her own value for $m$ if she wishes. We also ask the expert to propose a value for the control group mean $\mu_c$; we would expect appropriate values to be readily available from the literature. Without loss of generality, we will suppose in the following that $\mu_c=0$.

We then ask the expert to propose an interval on the outcome response scale $[k_1,k_2]$, that has meaningful interpretation in terms of patient outcomes. One possibility is to return to the notion of a minimum clinically relevent difference $\delta^*$, and the consider the interval $(-\infty, \delta^*]$. Hence (assuming $\mu_c=0$), any patient 
with an observation in this interval can be interpreted as having \textit{not} benefited from the treatment, even though the treatment is assumed to be effective `on average'. 

Finally, we define $\omega$ to be the proportion of patients who would have outcomes in the interval $(-\infty, \delta^*]$, and we ask the expert to consider her uncertainty about $\omega$. In summary, we are asking: ``Suppose the treatment does have the expected effect (on average). What proportion of patients might, nevertherless, \textit{not} achieve the desired response given the treatment?''

For certain choices of the interval $[k_1,k_2]$, there is a simple, montonic relationship between $\omega$ and $\sigma^2_t$. The interval will need to be in the form of one of the following: $(-\infty, k_2], [k_1, m], [m, k_2]$ or $[k_1,\infty)$. For example, for an interval of the form $(-\infty, k_2]$, we have
\begin{equation}
\sigma_t = \frac{k_2 - m}{\Phi^{-1}(\omega)}\label{eq:sigmaProportion}.
\end{equation}

As a simple example to visualise this, suppose we were to elicit an expert's beliefs about students' marks for an undergraduate statistics module, for a large population of students.  Suppose the marks are normally distributed with a mean of 60. Then there is a true proportion of students who will get marks between 60 and 70. If this expert, having been told that the mean is 60, is certain this proportion would be less that 0.45 and more than 0.25, this would imply she is certain $\sigma$ is between $6$ and $15$. This is illustrated in Figure \ref{fig:shaded-proportion}.       \begin{figure}[ht]
      \centering
\begin{tikzpicture}[x=1pt,y=1pt]
\definecolor{fillColor}{RGB}{255,255,255}
\path[use as bounding box,fill=fillColor,fill opacity=0.00] (0,0) rectangle (354.12,252.94);
\begin{scope}
\path[clip] ( 49.20, 61.20) rectangle (151.86,203.75);
\definecolor{drawColor}{RGB}{0,0,0}

\path[draw=drawColor,line width= 0.8pt,line join=round,line cap=round] ( 53.00, 66.48) --
	( 53.96, 66.48) --
	( 54.92, 66.48) --
	( 55.88, 66.48) --
	( 56.84, 66.48) --
	( 57.80, 66.48) --
	( 58.76, 66.48) --
	( 59.72, 66.48) --
	( 60.68, 66.48) --
	( 61.64, 66.48) --
	( 62.60, 66.48) --
	( 63.56, 66.48) --
	( 64.52, 66.48) --
	( 65.48, 66.48) --
	( 66.44, 66.48) --
	( 67.40, 66.48) --
	( 68.37, 66.48) --
	( 69.33, 66.49) --
	( 70.29, 66.49) --
	( 71.25, 66.50) --
	( 72.21, 66.52) --
	( 73.17, 66.55) --
	( 74.13, 66.59) --
	( 75.09, 66.67) --
	( 76.05, 66.78) --
	( 77.01, 66.95) --
	( 77.97, 67.21) --
	( 78.93, 67.59) --
	( 79.89, 68.14) --
	( 80.85, 68.91) --
	( 81.81, 69.97) --
	( 82.77, 71.40) --
	( 83.73, 73.30) --
	( 84.69, 75.77) --
	( 85.65, 78.89) --
	( 86.61, 82.78) --
	( 87.57, 87.49) --
	( 88.53, 93.07) --
	( 89.49, 99.54) --
	( 90.45,106.84) --
	( 91.41,114.87) --
	( 92.37,123.45) --
	( 93.33,132.34) --
	( 94.29,141.26) --
	( 95.25,149.86) --
	( 96.21,157.77) --
	( 97.17,164.64) --
	( 98.13,170.13) --
	( 99.09,173.96) --
	(100.05,175.93) --
	(101.01,175.93) --
	(101.97,173.96) --
	(102.93,170.13) --
	(103.89,164.64) --
	(104.85,157.77) --
	(105.81,149.86) --
	(106.77,141.26) --
	(107.73,132.34) --
	(108.69,123.45) --
	(109.65,114.87) --
	(110.61,106.84) --
	(111.57, 99.54) --
	(112.53, 93.07) --
	(113.49, 87.49) --
	(114.45, 82.78) --
	(115.41, 78.89) --
	(116.37, 75.77) --
	(117.33, 73.30) --
	(118.29, 71.40) --
	(119.25, 69.97) --
	(120.21, 68.91) --
	(121.17, 68.14) --
	(122.13, 67.59) --
	(123.09, 67.21) --
	(124.05, 66.95) --
	(125.02, 66.78) --
	(125.98, 66.67) --
	(126.94, 66.59) --
	(127.90, 66.55) --
	(128.86, 66.52) --
	(129.82, 66.50) --
	(130.78, 66.49) --
	(131.74, 66.49) --
	(132.70, 66.48) --
	(133.66, 66.48) --
	(134.62, 66.48) --
	(135.58, 66.48) --
	(136.54, 66.48) --
	(137.50, 66.48) --
	(138.46, 66.48) --
	(139.42, 66.48) --
	(140.38, 66.48) --
	(141.34, 66.48) --
	(142.30, 66.48) --
	(143.26, 66.48) --
	(144.22, 66.48) --
	(145.18, 66.48) --
	(146.14, 66.48) --
	(147.10, 66.48) --
	(148.06, 66.48);
\end{scope}
\begin{scope}
\path[clip] (  0.00,  0.00) rectangle (354.12,252.94);
\definecolor{drawColor}{RGB}{0,0,0}

\path[draw=drawColor,line width= 0.4pt,line join=round,line cap=round] ( 53.00, 61.20) -- (148.06, 61.20);

\path[draw=drawColor,line width= 0.4pt,line join=round,line cap=round] ( 53.00, 61.20) -- ( 53.00, 55.20);

\path[draw=drawColor,line width= 0.4pt,line join=round,line cap=round] ( 76.77, 61.20) -- ( 76.77, 55.20);

\path[draw=drawColor,line width= 0.4pt,line join=round,line cap=round] (100.53, 61.20) -- (100.53, 55.20);

\path[draw=drawColor,line width= 0.4pt,line join=round,line cap=round] (124.29, 61.20) -- (124.29, 55.20);

\path[draw=drawColor,line width= 0.4pt,line join=round,line cap=round] (148.06, 61.20) -- (148.06, 55.20);

\node[text=drawColor,anchor=base,inner sep=0pt, outer sep=0pt, scale=  1.00] at ( 53.00, 39.60) {20};

\node[text=drawColor,anchor=base,inner sep=0pt, outer sep=0pt, scale=  1.00] at ( 76.77, 39.60) {40};

\node[text=drawColor,anchor=base,inner sep=0pt, outer sep=0pt, scale=  1.00] at (100.53, 39.60) {60};

\node[text=drawColor,anchor=base,inner sep=0pt, outer sep=0pt, scale=  1.00] at (124.29, 39.60) {80};

\node[text=drawColor,anchor=base,inner sep=0pt, outer sep=0pt, scale=  1.00] at (148.06, 39.60) {100};

\path[draw=drawColor,line width= 0.4pt,line join=round,line cap=round] ( 49.20, 66.48) -- ( 49.20,198.47);

\path[draw=drawColor,line width= 0.4pt,line join=round,line cap=round] ( 49.20, 66.48) -- ( 43.20, 66.48);

\path[draw=drawColor,line width= 0.4pt,line join=round,line cap=round] ( 49.20, 99.48) -- ( 43.20, 99.48);

\path[draw=drawColor,line width= 0.4pt,line join=round,line cap=round] ( 49.20,132.47) -- ( 43.20,132.47);

\path[draw=drawColor,line width= 0.4pt,line join=round,line cap=round] ( 49.20,165.47) -- ( 43.20,165.47);

\path[draw=drawColor,line width= 0.4pt,line join=round,line cap=round] ( 49.20,198.47) -- ( 43.20,198.47);

\node[text=drawColor,rotate= 90.00,anchor=base,inner sep=0pt, outer sep=0pt, scale=  1.00] at ( 34.80, 66.48) {0.00};

\node[text=drawColor,rotate= 90.00,anchor=base,inner sep=0pt, outer sep=0pt, scale=  1.00] at ( 34.80, 99.48) {0.02};

\node[text=drawColor,rotate= 90.00,anchor=base,inner sep=0pt, outer sep=0pt, scale=  1.00] at ( 34.80,132.47) {0.04};

\node[text=drawColor,rotate= 90.00,anchor=base,inner sep=0pt, outer sep=0pt, scale=  1.00] at ( 34.80,165.47) {0.06};

\node[text=drawColor,rotate= 90.00,anchor=base,inner sep=0pt, outer sep=0pt, scale=  1.00] at ( 34.80,198.47) {0.08};

\path[draw=drawColor,line width= 0.4pt,line join=round,line cap=round] ( 49.20, 61.20) --
	(151.86, 61.20) --
	(151.86,203.75) --
	( 49.20,203.75) --
	( 49.20, 61.20);
\end{scope}
\begin{scope}
\path[clip] (  0.00,  0.00) rectangle (177.06,252.94);
\definecolor{drawColor}{RGB}{0,0,0}

\node[text=drawColor,anchor=base,inner sep=0pt, outer sep=0pt, scale=  1.00] at (100.53, 15.60) {$x$};

\node[text=drawColor,rotate= 90.00,anchor=base,inner sep=0pt, outer sep=0pt, scale=  1.00] at ( 10.80,132.47) {$f_X(x|\mu=60, \sigma^2)$};
\end{scope}
\begin{scope}
\path[clip] (  0.00,  0.00) rectangle (354.12,252.94);
\definecolor{drawColor}{RGB}{0,0,0}

\node[text=drawColor,anchor=base,inner sep=0pt, outer sep=0pt, scale=  1.00] at (100.53,212.14) {$\sigma=$ 6};
\end{scope}
\begin{scope}
\path[clip] ( 49.20, 61.20) rectangle (151.86,203.75);
\definecolor{drawColor}{RGB}{0,0,0}
\definecolor{fillColor}{RGB}{190,190,190}

\path[draw=drawColor,line width= 0.4pt,line join=round,line cap=round,fill=fillColor] (100.53, 66.48) --
	(100.53,176.18) --
	(100.65,176.16) --
	(100.77,176.11) --
	(100.89,176.04) --
	(101.01,175.93) --
	(101.13,175.79) --
	(101.25,175.62) --
	(101.37,175.42) --
	(101.49,175.19) --
	(101.61,174.93) --
	(101.73,174.63) --
	(101.85,174.31) --
	(101.97,173.96) --
	(102.09,173.58) --
	(102.21,173.17) --
	(102.33,172.73) --
	(102.45,172.27) --
	(102.57,171.78) --
	(102.69,171.25) --
	(102.81,170.71) --
	(102.93,170.13) --
	(103.05,169.53) --
	(103.17,168.91) --
	(103.29,168.25) --
	(103.41,167.58) --
	(103.53,166.88) --
	(103.65,166.16) --
	(103.77,165.41) --
	(103.89,164.64) --
	(104.01,163.85) --
	(104.13,163.04) --
	(104.25,162.21) --
	(104.37,161.36) --
	(104.49,160.49) --
	(104.61,159.60) --
	(104.73,158.70) --
	(104.85,157.77) --
	(104.97,156.83) --
	(105.09,155.88) --
	(105.21,154.91) --
	(105.33,153.92) --
	(105.45,152.92) --
	(105.57,151.91) --
	(105.69,150.89) --
	(105.81,149.86) --
	(105.93,148.81) --
	(106.05,147.76) --
	(106.17,146.69) --
	(106.29,145.62) --
	(106.41,144.54) --
	(106.53,143.45) --
	(106.65,142.36) --
	(106.77,141.26) --
	(106.89,140.15) --
	(107.01,139.05) --
	(107.13,137.93) --
	(107.25,136.82) --
	(107.37,135.70) --
	(107.49,134.58) --
	(107.61,133.46) --
	(107.73,132.34) --
	(107.85,131.22) --
	(107.97,130.10) --
	(108.09,128.99) --
	(108.21,127.87) --
	(108.33,126.76) --
	(108.45,125.65) --
	(108.57,124.55) --
	(108.69,123.45) --
	(108.81,122.35) --
	(108.93,121.26) --
	(109.05,120.18) --
	(109.17,119.10) --
	(109.29,118.03) --
	(109.41,116.97) --
	(109.53,115.91) --
	(109.65,114.87) --
	(109.77,113.83) --
	(109.89,112.80) --
	(110.01,111.78) --
	(110.13,110.77) --
	(110.25,109.77) --
	(110.37,108.78) --
	(110.49,107.81) --
	(110.61,106.84) --
	(110.73,105.88) --
	(110.85,104.94) --
	(110.97,104.01) --
	(111.09,103.09) --
	(111.21,102.18) --
	(111.33,101.29) --
	(111.45,100.41) --
	(111.57, 99.54) --
	(111.69, 98.68) --
	(111.81, 97.84) --
	(111.93, 97.01) --
	(112.05, 96.20) --
	(112.17, 95.40) --
	(112.29, 94.61) --
	(112.41, 93.83) --
	(112.41, 66.48) --
	cycle;
\end{scope}
\begin{scope}
\path[clip] (226.26, 61.20) rectangle (328.92,203.75);
\definecolor{drawColor}{RGB}{0,0,0}

\path[draw=drawColor,line width= 0.8pt,line join=round,line cap=round] (230.06, 67.73) --
	(231.02, 67.92) --
	(231.98, 68.14) --
	(232.94, 68.38) --
	(233.90, 68.66) --
	(234.86, 68.96) --
	(235.82, 69.30) --
	(236.78, 69.67) --
	(237.75, 70.08) --
	(238.71, 70.54) --
	(239.67, 71.04) --
	(240.63, 71.59) --
	(241.59, 72.18) --
	(242.55, 72.83) --
	(243.51, 73.53) --
	(244.47, 74.28) --
	(245.43, 75.09) --
	(246.39, 75.95) --
	(247.35, 76.88) --
	(248.31, 77.86) --
	(249.27, 78.89) --
	(250.23, 79.98) --
	(251.19, 81.12) --
	(252.15, 82.32) --
	(253.11, 83.56) --
	(254.07, 84.84) --
	(255.03, 86.17) --
	(255.99, 87.53) --
	(256.95, 88.92) --
	(257.91, 90.33) --
	(258.87, 91.75) --
	(259.83, 93.18) --
	(260.79, 94.62) --
	(261.75, 96.04) --
	(262.71, 97.44) --
	(263.67, 98.82) --
	(264.63,100.16) --
	(265.59,101.46) --
	(266.55,102.70) --
	(267.51,103.87) --
	(268.47,104.97) --
	(269.43,105.99) --
	(270.39,106.92) --
	(271.35,107.75) --
	(272.31,108.47) --
	(273.27,109.09) --
	(274.23,109.59) --
	(275.19,109.96) --
	(276.15,110.22) --
	(277.11,110.34) --
	(278.07,110.34) --
	(279.03,110.22) --
	(279.99,109.96) --
	(280.95,109.59) --
	(281.91,109.09) --
	(282.87,108.47) --
	(283.83,107.75) --
	(284.79,106.92) --
	(285.75,105.99) --
	(286.71,104.97) --
	(287.67,103.87) --
	(288.63,102.70) --
	(289.59,101.46) --
	(290.55,100.16) --
	(291.51, 98.82) --
	(292.47, 97.44) --
	(293.44, 96.04) --
	(294.40, 94.62) --
	(295.36, 93.18) --
	(296.32, 91.75) --
	(297.28, 90.33) --
	(298.24, 88.92) --
	(299.20, 87.53) --
	(300.16, 86.17) --
	(301.12, 84.84) --
	(302.08, 83.56) --
	(303.04, 82.32) --
	(304.00, 81.12) --
	(304.96, 79.98) --
	(305.92, 78.89) --
	(306.88, 77.86) --
	(307.84, 76.88) --
	(308.80, 75.95) --
	(309.76, 75.09) --
	(310.72, 74.28) --
	(311.68, 73.53) --
	(312.64, 72.83) --
	(313.60, 72.18) --
	(314.56, 71.59) --
	(315.52, 71.04) --
	(316.48, 70.54) --
	(317.44, 70.08) --
	(318.40, 69.67) --
	(319.36, 69.30) --
	(320.32, 68.96) --
	(321.28, 68.66) --
	(322.24, 68.38) --
	(323.20, 68.14) --
	(324.16, 67.92) --
	(325.12, 67.73);
\end{scope}
\begin{scope}
\path[clip] (  0.00,  0.00) rectangle (354.12,252.94);
\definecolor{drawColor}{RGB}{0,0,0}

\path[draw=drawColor,line width= 0.4pt,line join=round,line cap=round] (230.06, 61.20) -- (325.12, 61.20);

\path[draw=drawColor,line width= 0.4pt,line join=round,line cap=round] (230.06, 61.20) -- (230.06, 55.20);

\path[draw=drawColor,line width= 0.4pt,line join=round,line cap=round] (253.83, 61.20) -- (253.83, 55.20);

\path[draw=drawColor,line width= 0.4pt,line join=round,line cap=round] (277.59, 61.20) -- (277.59, 55.20);

\path[draw=drawColor,line width= 0.4pt,line join=round,line cap=round] (301.36, 61.20) -- (301.36, 55.20);

\path[draw=drawColor,line width= 0.4pt,line join=round,line cap=round] (325.12, 61.20) -- (325.12, 55.20);

\node[text=drawColor,anchor=base,inner sep=0pt, outer sep=0pt, scale=  1.00] at (230.06, 39.60) {20};

\node[text=drawColor,anchor=base,inner sep=0pt, outer sep=0pt, scale=  1.00] at (253.83, 39.60) {40};

\node[text=drawColor,anchor=base,inner sep=0pt, outer sep=0pt, scale=  1.00] at (277.59, 39.60) {60};

\node[text=drawColor,anchor=base,inner sep=0pt, outer sep=0pt, scale=  1.00] at (301.36, 39.60) {80};

\node[text=drawColor,anchor=base,inner sep=0pt, outer sep=0pt, scale=  1.00] at (325.12, 39.60) {100};

\path[draw=drawColor,line width= 0.4pt,line join=round,line cap=round] (226.26, 66.48) -- (226.26,198.47);

\path[draw=drawColor,line width= 0.4pt,line join=round,line cap=round] (226.26, 66.48) -- (220.26, 66.48);

\path[draw=drawColor,line width= 0.4pt,line join=round,line cap=round] (226.26, 99.48) -- (220.26, 99.48);

\path[draw=drawColor,line width= 0.4pt,line join=round,line cap=round] (226.26,132.47) -- (220.26,132.47);

\path[draw=drawColor,line width= 0.4pt,line join=round,line cap=round] (226.26,165.47) -- (220.26,165.47);

\path[draw=drawColor,line width= 0.4pt,line join=round,line cap=round] (226.26,198.47) -- (220.26,198.47);

\node[text=drawColor,rotate= 90.00,anchor=base,inner sep=0pt, outer sep=0pt, scale=  1.00] at (211.86, 66.48) {0.00};

\node[text=drawColor,rotate= 90.00,anchor=base,inner sep=0pt, outer sep=0pt, scale=  1.00] at (211.86, 99.48) {0.02};

\node[text=drawColor,rotate= 90.00,anchor=base,inner sep=0pt, outer sep=0pt, scale=  1.00] at (211.86,132.47) {0.04};

\node[text=drawColor,rotate= 90.00,anchor=base,inner sep=0pt, outer sep=0pt, scale=  1.00] at (211.86,165.47) {0.06};

\node[text=drawColor,rotate= 90.00,anchor=base,inner sep=0pt, outer sep=0pt, scale=  1.00] at (211.86,198.47) {0.08};

\path[draw=drawColor,line width= 0.4pt,line join=round,line cap=round] (226.26, 61.20) --
	(328.92, 61.20) --
	(328.92,203.75) --
	(226.26,203.75) --
	(226.26, 61.20);
\end{scope}
\begin{scope}
\path[clip] (177.06,  0.00) rectangle (354.12,252.94);
\definecolor{drawColor}{RGB}{0,0,0}

\node[text=drawColor,anchor=base,inner sep=0pt, outer sep=0pt, scale=  1.00] at (277.59, 15.60) {$x$};

\node[text=drawColor,rotate= 90.00,anchor=base,inner sep=0pt, outer sep=0pt, scale=  1.00] at (187.86,132.47) {$f_X(x|\mu=60, \sigma^2)$};
\end{scope}
\begin{scope}
\path[clip] (  0.00,  0.00) rectangle (354.12,252.94);
\definecolor{drawColor}{RGB}{0,0,0}

\node[text=drawColor,anchor=base,inner sep=0pt, outer sep=0pt, scale=  1.00] at (277.59,212.14) {$\sigma=$ 15};
\end{scope}
\begin{scope}
\path[clip] (226.26, 61.20) rectangle (328.92,203.75);
\definecolor{drawColor}{RGB}{0,0,0}
\definecolor{fillColor}{RGB}{190,190,190}

\path[draw=drawColor,line width= 0.4pt,line join=round,line cap=round,fill=fillColor] (277.59, 66.48) --
	(277.59,110.36) --
	(277.71,110.36) --
	(277.83,110.35) --
	(277.95,110.35) --
	(278.07,110.34) --
	(278.19,110.33) --
	(278.31,110.32) --
	(278.43,110.31) --
	(278.55,110.29) --
	(278.67,110.28) --
	(278.79,110.26) --
	(278.91,110.24) --
	(279.03,110.22) --
	(279.15,110.19) --
	(279.27,110.16) --
	(279.39,110.14) --
	(279.51,110.10) --
	(279.63,110.07) --
	(279.75,110.04) --
	(279.87,110.00) --
	(279.99,109.96) --
	(280.11,109.92) --
	(280.23,109.88) --
	(280.35,109.84) --
	(280.47,109.79) --
	(280.59,109.74) --
	(280.71,109.69) --
	(280.83,109.64) --
	(280.95,109.59) --
	(281.07,109.53) --
	(281.19,109.47) --
	(281.31,109.41) --
	(281.43,109.35) --
	(281.55,109.29) --
	(281.67,109.22) --
	(281.79,109.16) --
	(281.91,109.09) --
	(282.03,109.02) --
	(282.15,108.95) --
	(282.27,108.87) --
	(282.39,108.80) --
	(282.51,108.72) --
	(282.63,108.64) --
	(282.75,108.56) --
	(282.87,108.47) --
	(282.99,108.39) --
	(283.11,108.30) --
	(283.23,108.21) --
	(283.35,108.13) --
	(283.47,108.03) --
	(283.59,107.94) --
	(283.71,107.85) --
	(283.83,107.75) --
	(283.95,107.65) --
	(284.07,107.55) --
	(284.19,107.45) --
	(284.31,107.35) --
	(284.43,107.24) --
	(284.55,107.14) --
	(284.67,107.03) --
	(284.79,106.92) --
	(284.91,106.81) --
	(285.03,106.70) --
	(285.15,106.58) --
	(285.27,106.47) --
	(285.39,106.35) --
	(285.51,106.23) --
	(285.63,106.11) --
	(285.75,105.99) --
	(285.87,105.87) --
	(285.99,105.74) --
	(286.11,105.62) --
	(286.23,105.49) --
	(286.35,105.36) --
	(286.47,105.24) --
	(286.59,105.10) --
	(286.71,104.97) --
	(286.83,104.84) --
	(286.95,104.70) --
	(287.07,104.57) --
	(287.19,104.43) --
	(287.31,104.29) --
	(287.43,104.15) --
	(287.55,104.01) --
	(287.67,103.87) --
	(287.79,103.73) --
	(287.91,103.58) --
	(288.03,103.44) --
	(288.15,103.29) --
	(288.27,103.15) --
	(288.39,103.00) --
	(288.51,102.85) --
	(288.63,102.70) --
	(288.75,102.54) --
	(288.87,102.39) --
	(288.99,102.24) --
	(289.11,102.08) --
	(289.23,101.93) --
	(289.35,101.77) --
	(289.47,101.62) --
	(289.47, 66.48) --
	cycle;
\end{scope}
\end{tikzpicture}

       \caption{Density plots for exam marks assumed to be normally distributed with mean 60. The grey area represents the true proportion of students who get marks between 60 and 70. If the expert is certain this proportion is between 0.25 and 0.45, she is certain $\sigma$ is between 6 and 15.}\label{fig:shaded-proportion}
    \end{figure}

Probability judgements about $\omega$ can be converted to probability judgements about $\sigma_t$, and so we can use the same approach as described in Section \ref{sec:uniElicitation}. However, given the somewhat abstract nature of $\omega$, we would suggest asking the expert for `lower' and `upper' bounds, which we would then interpret as 5th and 95th percentiles, and denote by $\omega_{0.05}$ and $\omega_{0.95}$ respectively.

It may help the expert to explicitly consider two intervals, for example $(-\infty, \delta^*]$ and  $[\delta^*, m]$ and then consider how the population is distributed between these two intervals. For example, she might judge a split of 2\%-48\% across the two intervals highly unlikely, which can help prompt judgements of more plausible allocations (though the facilitator should remain cautious of anchoring effects).

Given $\omega_{0.05}$ and $\omega_{0.95}$, we can infer the corresponding quantiles of her distribution for the variance, which we denote by ${\sigma}^2_{t, (0.95)}$ and ${\sigma}^2_{t, (0.05)}$, using equation (\ref{eq:sigmaProportion}). The facilitator now chooses a parametric family of distributions, and obtains the parameter values $\theta$ within that family by minimising (numerically)
$$
\left(F_{\theta}({\sigma}^2_{(0.05)}) - 0.05\right)^2 + \left(F_{\theta}({\sigma}^2_{(0.95)}) - 0.95\right)^2
$$
where $F_{\theta}(.)$ is the cumulative distribution function from the chosen family, with parameter values $\theta$. We suggest fitting a distribution to the precision $\sigma^{-2}$, and choosing either a log-normal distribution or a gamma distribution. 

The log-normal and gamma distributions can fit the two judgements precisely, so to guide the choice of distribution, we suggest presenting the implied distribution of the standard deviation $\sigma$, indicating 5th and 95th percentiles; differences between the two fits may be more apparent in the tails.

\subsection{Distribution for the control group variance}

There are a number of options that could be used for the control group variance $\sigma^2_c$:
\begin{enumerate}
\item use a point estimate or a distribution based on historical data;
\item assume that $\sigma^2_c = \sigma^2_t$;
\item assume that $\sigma^2_c$ and $\sigma^2_t$ are independent and identically distributed;
\item elicit a separate distribution for $\sigma^2_c$.
\end{enumerate}
(A fifth option could involve a hierarchical model for $\sigma^2_c$ and $\sigma^2_t$, so that learning about one could update beliefs about the other, but this would make the elicitation task considerably more difficult).

We think the first option would be the most commonly used in practice. Note that, if a mixture prior has been used for $\delta$, one might then also consider a mixture prior for $\sigma^2_t$: if $\delta=0$, and assuming that in that case, the treatment is no different to a placebo, one might also suppose $\sigma_t^2 = \sigma_c^2$.
\section{Computing assurances}

Given the elicited prior, we can now compute the assurance for any choice of sample sizes, using the following algorithm.

\noindent\textbf{Algorithm 1: estimating an assurance}\\
Inputs: sample sizes $n_t$ and $n_c$, the elicited prior $\pi(\delta, \sigma^2_t, \sigma^2_c)$, and the number of iterations $N$. \\
For $i=1,\ldots,N$:
\begin{enumerate}
\item sample $\delta_i,\sigma^2_{t, i}$ and $\sigma^2_{c,i}$ from $\pi(\delta, \sigma^2_t, \sigma^2_c)$;
\item sample $x_{1,i},\ldots,x_{n_t,i}$ from $N(\delta_i, \sigma^2_{t,i})$ and $y_{1,i},\ldots,y_{n_c,i}$ from $N(0, \sigma^2_{c,i})$;
\item calculate $\bar{x}_i, s^2_{x,i}$ as the sample mean and sample variance of $x_{1,i},\ldots,x_{n_t,i}$, and $\bar{y}_i, s^2_{y,i}$ as the sample mean and sample variance of $y_{1,i},\ldots,y_{n_c,i}$;
\item calculate the test statistic $T_i$ and degrees of freedom $\nu_i$:
\begin{eqnarray*}
T_i &=& \frac{\bar{x}_i - \bar{y}_i}{\sqrt{\frac{s^2_{x,i}}{n_{t}} + \frac{s^2_{y,i}}{n_{c}}}},\\
\nu_i &=& \frac {\left(\frac{s^2_{x,i}}{n_t} + \frac{s^2_{y,i}}{n_c}\right)^2}
            {\frac{(s^2_{x,i}/n_t)^2}{n_t-1} + \frac{(s^2_{y,i}/n_c)^2}{n_c-1}};
\end{eqnarray*}
\item define $R_i = 1$ if $T_i > t_{0.025, \nu_i}$ and 0 otherwise, with $t_{0.025, \nu_i}$ the 97.5th percentile of the Student-$t$ distribution with $\nu_i$ degrees of freedom. (We assume here that we require $\bar{x}_i > \bar{y}_i$ for the treatment effect to be beneficial).
\end{enumerate}
The assurance is then estimated as
$$
\hat{P}(R) = \frac{1}{N}\sum_{i=1}^N R_i.
$$

\subsection{Example \label{sec:Example}}
We illustrate the method with a (fictitious) example based on Example 3 from \cite{O'Hagan:2005} (with slight modifications to their numerical values). In their example, they consider a Phase 2 superiority trial to assess the effect of a new drug in reducing C-reactive protein (CRP) in patients with rheumatoid arthritis. Their outcome variable is a patient's reduction in CRP after 4 weeks relative to baseline, and the analysis to be performed is a two-sided test of superiority at the 5\% significance level. They considered power calculations assuming $\delta=0.2$, so we will suppose that 0.2 is the minimum clinically relevant treatment effect.

\subsubsection{The prior distribution for the treatment effect}

We suppose that the expert judges a non-zero probability that the treatment will have no effect, but we will consider two scenarios: in the first, the expert judges $Pr(\delta=0) = 0.5$, and in the second, the expert is more optimistic with $Pr(\delta=0) = 0.1$. In both scenarios,  conditional on $\delta\neq 0$, we suppose that the expert provides three quartiles from her distribution for $\delta$: 0.25, 0.4 and 0.55. A normal distribution with mean 0.4 and standard deviation 0.22 is fitted to these judgements. 

\subsubsection{The prior distribution for the variances}

The expert is asked to assume that (a) the treatment is effective, with $\delta$ equal to 0.4; (b) in the control group, the mean reduction in CRP would be 0; (c) individual patients with reduction from baseline of 0.2 or less would not be judged to have received a clinically meaningful benefit. She is then asked to consider, under these assumptions, what proportion $\omega$ of patients in the treatment group would not benefit: the proportion of patients with reductions less than 0.2.

We suppose she judges that this proportion will be between 20\% and 40\%, which we judge to be the 5th and 95th percentiles of her distribution for $\omega$. These correspond to 5th and 95th percentiles of her distribution for $\sigma_t$ of 0.24 and 0.8 respectively. We fit a gamma distribution to her precision $\sigma_t^{-2}$ with shape parameter 2.27 and rate parameter 0.29.

Finally, we suppose that the expert judges that the same distribution will be appropriate for $\sigma^2_c$. In the case that $\delta\neq 0$, she judges $\sigma^2_c$ and $\sigma^2_t$ to be independent, and if $\delta= 0$ then she judges $\sigma^2_c=\sigma^2_t$.

\subsubsection{Estimating assurances}

Assurances are estimated using Algorithm 1. We illustrate the results that would be presented to the trial planners in Table \ref{assuranceTable1}, assuming equal numbers in the treatment and control arms.

\begin{table}[H]
\begin{center}
\begin{tabular}{r|c|c|c|c|c}
Sample size per arm & 10 & 20 & 50 & 100 & 1000\\ \hline
$Pr(\delta=0) = 0.5$ & 0.28  & 0.36 &  0.42 &  0.45 &  0.49 \\
$Pr(\delta=0) = 0.1$ & 0.48 & 0.62 &  0.74 &  0.79 &  0.86
\end{tabular}
\caption{Estimated assurances for the two prior elicited prior distributions. There is the same `diminishing return' from increasing the sample size as one would see in a power curve, but the assurance converges (approximately) to the prior probability that the treatment is effective.} \label{assuranceTable1}
\end{center}
\end{table}

The assurance will have an upper bound close to the prior probability that the treatment is effective (but not necessarily equal to this probability, given the possibility of Type I errors). It can be informative to consider the assurance as a fraction of this prior probability. We define a ``scaled assurance'' as $Pr(R) / Pr(\delta > 0)$, and plot this in Figure \ref{scaledAssurance}. We can see that for the two scenarios, the scaled assurances are fairly similar. This may be useful for sample size planning where it is difficult for the expert to specify a precise value for $Pr(\delta = 0)$: the scaled assurance `levels off' at similar sample sizes. 

We also plot scaled assurances where $\sigma_t$ and $\sigma_c$ are held equal and fixed at one of the two elicited values: $\sigma_{t,0.05} = 0.24$ and $\sigma_{t,0.95} = 0.8$. Again, within each scenario, the scaled assurances are similar, though uncertainty about $\sigma_t$ clearly matters in this example.

\begin{figure}[H]
\begin{tikzpicture}[x=1pt,y=1pt]
\definecolor{fillColor}{RGB}{255,255,255}
\path[use as bounding box,fill=fillColor,fill opacity=0.00] (0,0) rectangle (430.01,303.53);
\begin{scope}
\path[clip] ( 49.20, 61.20) rectangle (404.81,254.33);
\definecolor{drawColor}{RGB}{0,0,0}

\path[draw=drawColor,line width= 0.4pt,line join=round,line cap=round] ( 62.37,132.92) --
	( 79.70,170.93) --
	( 97.03,189.24) --
	(114.36,200.28) --
	(131.69,208.82) --
	(149.02,213.11) --
	(166.35,217.81) --
	(183.68,220.76) --
	(201.01,223.39) --
	(218.34,225.95) --
	(235.67,227.93) --
	(253.00,228.21) --
	(270.33,230.56) --
	(287.66,231.03) --
	(304.99,232.28) --
	(322.32,233.75) --
	(339.65,234.06) --
	(356.98,235.04) --
	(374.31,235.61) --
	(391.64,236.12);
\end{scope}
\begin{scope}
\path[clip] (  0.00,  0.00) rectangle (430.01,303.53);
\definecolor{drawColor}{RGB}{0,0,0}

\path[draw=drawColor,line width= 0.4pt,line join=round,line cap=round] (114.36, 61.20) -- (391.64, 61.20);

\path[draw=drawColor,line width= 0.4pt,line join=round,line cap=round] (114.36, 61.20) -- (114.36, 55.20);

\path[draw=drawColor,line width= 0.4pt,line join=round,line cap=round] (183.68, 61.20) -- (183.68, 55.20);

\path[draw=drawColor,line width= 0.4pt,line join=round,line cap=round] (253.00, 61.20) -- (253.00, 55.20);

\path[draw=drawColor,line width= 0.4pt,line join=round,line cap=round] (322.32, 61.20) -- (322.32, 55.20);

\path[draw=drawColor,line width= 0.4pt,line join=round,line cap=round] (391.64, 61.20) -- (391.64, 55.20);

\node[text=drawColor,anchor=base,inner sep=0pt, outer sep=0pt, scale=  1.00] at (114.36, 39.60) {20};

\node[text=drawColor,anchor=base,inner sep=0pt, outer sep=0pt, scale=  1.00] at (183.68, 39.60) {40};

\node[text=drawColor,anchor=base,inner sep=0pt, outer sep=0pt, scale=  1.00] at (253.00, 39.60) {60};

\node[text=drawColor,anchor=base,inner sep=0pt, outer sep=0pt, scale=  1.00] at (322.32, 39.60) {80};

\node[text=drawColor,anchor=base,inner sep=0pt, outer sep=0pt, scale=  1.00] at (391.64, 39.60) {100};

\path[draw=drawColor,line width= 0.4pt,line join=round,line cap=round] ( 49.20, 68.35) -- ( 49.20,247.18);

\path[draw=drawColor,line width= 0.4pt,line join=round,line cap=round] ( 49.20, 68.35) -- ( 43.20, 68.35);

\path[draw=drawColor,line width= 0.4pt,line join=round,line cap=round] ( 49.20,104.12) -- ( 43.20,104.12);

\path[draw=drawColor,line width= 0.4pt,line join=round,line cap=round] ( 49.20,139.88) -- ( 43.20,139.88);

\path[draw=drawColor,line width= 0.4pt,line join=round,line cap=round] ( 49.20,175.65) -- ( 43.20,175.65);

\path[draw=drawColor,line width= 0.4pt,line join=round,line cap=round] ( 49.20,211.42) -- ( 43.20,211.42);

\path[draw=drawColor,line width= 0.4pt,line join=round,line cap=round] ( 49.20,247.18) -- ( 43.20,247.18);

\node[text=drawColor,rotate= 90.00,anchor=base,inner sep=0pt, outer sep=0pt, scale=  1.00] at ( 34.80, 68.35) {0.0};

\node[text=drawColor,rotate= 90.00,anchor=base,inner sep=0pt, outer sep=0pt, scale=  1.00] at ( 34.80,104.12) {0.2};

\node[text=drawColor,rotate= 90.00,anchor=base,inner sep=0pt, outer sep=0pt, scale=  1.00] at ( 34.80,139.88) {0.4};

\node[text=drawColor,rotate= 90.00,anchor=base,inner sep=0pt, outer sep=0pt, scale=  1.00] at ( 34.80,175.65) {0.6};

\node[text=drawColor,rotate= 90.00,anchor=base,inner sep=0pt, outer sep=0pt, scale=  1.00] at ( 34.80,211.42) {0.8};

\node[text=drawColor,rotate= 90.00,anchor=base,inner sep=0pt, outer sep=0pt, scale=  1.00] at ( 34.80,247.18) {1.0};

\path[draw=drawColor,line width= 0.4pt,line join=round,line cap=round] ( 49.20, 61.20) --
	(404.81, 61.20) --
	(404.81,254.33) --
	( 49.20,254.33) --
	( 49.20, 61.20);
\end{scope}
\begin{scope}
\path[clip] (  0.00,  0.00) rectangle (430.01,303.53);
\definecolor{drawColor}{RGB}{0,0,0}

\node[text=drawColor,anchor=base,inner sep=0pt, outer sep=0pt, scale=  1.00] at (227.00, 15.60) {sample size per arm};

\node[text=drawColor,rotate= 90.00,anchor=base,inner sep=0pt, outer sep=0pt, scale=  1.00] at ( 10.80,157.77) {scaled assurance};
\end{scope}
\begin{scope}
\path[clip] ( 49.20, 61.20) rectangle (404.81,254.33);
\definecolor{drawColor}{RGB}{255,0,0}

\path[draw=drawColor,line width= 0.4pt,dash pattern=on 4pt off 4pt ,line join=round,line cap=round] ( 62.37,129.90) --
	( 79.70,166.73) --
	( 97.03,185.63) --
	(114.36,196.59) --
	(131.69,203.94) --
	(149.02,209.26) --
	(166.35,212.85) --
	(183.68,216.02) --
	(201.01,219.07) --
	(218.34,221.48) --
	(235.67,223.54) --
	(253.00,224.83) --
	(270.33,226.09) --
	(287.66,227.36) --
	(304.99,228.44) --
	(322.32,229.28) --
	(339.65,230.26) --
	(356.98,230.86) --
	(374.31,231.21) --
	(391.64,231.95);

\path[draw=drawColor,line width= 0.4pt,dash pattern=on 4pt off 4pt ,line join=round,line cap=round] ( 62.37, 89.41) --
	( 79.70,108.89) --
	( 97.03,125.06) --
	(114.36,138.24) --
	(131.69,149.63) --
	(149.02,158.98) --
	(166.35,166.98) --
	(183.68,173.52) --
	(201.01,178.07) --
	(218.34,183.90) --
	(235.67,187.66) --
	(253.00,191.18) --
	(270.33,194.78) --
	(287.66,197.91) --
	(304.99,200.38) --
	(322.32,203.08) --
	(339.65,204.76) --
	(356.98,206.48) --
	(374.31,208.18) --
	(391.64,209.58);

\path[draw=drawColor,line width= 0.4pt,dash pattern=on 4pt off 4pt ,line join=round,line cap=round] ( 62.37,174.18) --
	( 79.70,209.98) --
	( 97.03,221.16) --
	(114.36,227.28) --
	(131.69,230.35) --
	(149.02,232.90) --
	(166.35,234.48) --
	(183.68,235.89) --
	(201.01,237.03) --
	(218.34,237.47) --
	(235.67,238.24) --
	(253.00,239.00) --
	(270.33,239.46) --
	(287.66,239.74) --
	(304.99,240.06) --
	(322.32,240.78) --
	(339.65,240.82) --
	(356.98,240.83) --
	(374.31,241.34) --
	(391.64,241.44);
\definecolor{drawColor}{RGB}{0,0,0}

\path[draw=drawColor,line width= 0.4pt,line join=round,line cap=round] ( 62.37, 92.74) --
	( 79.70,112.29) --
	( 97.03,128.76) --
	(114.36,141.99) --
	(131.69,153.40) --
	(149.02,162.69) --
	(166.35,170.39) --
	(183.68,177.75) --
	(201.01,182.88) --
	(218.34,187.82) --
	(235.67,192.12) --
	(253.00,195.24) --
	(270.33,198.87) --
	(287.66,201.08) --
	(304.99,204.42) --
	(322.32,206.93) --
	(339.65,208.60) --
	(356.98,210.93) --
	(374.31,212.80) --
	(391.64,213.87);

\path[draw=drawColor,line width= 0.4pt,line join=round,line cap=round] ( 62.37,177.46) --
	( 79.70,214.00) --
	( 97.03,225.58) --
	(114.36,231.12) --
	(131.69,234.47) --
	(149.02,236.80) --
	(166.35,239.02) --
	(183.68,239.79) --
	(201.01,240.26) --
	(218.34,242.03) --
	(235.67,242.62) --
	(253.00,242.88) --
	(270.33,243.60) --
	(287.66,243.78) --
	(304.99,244.02) --
	(322.32,244.70) --
	(339.65,245.21) --
	(356.98,244.73) --
	(374.31,245.24) --
	(391.64,246.11);

\path[draw=drawColor,line width= 0.4pt,line join=round,line cap=round] (284.67, 97.20) rectangle (404.81, 61.20);

\path[draw=drawColor,line width= 0.4pt,line join=round,line cap=round] (293.67, 85.20) -- (311.67, 85.20);
\definecolor{drawColor}{RGB}{255,0,0}

\path[draw=drawColor,line width= 0.4pt,dash pattern=on 4pt off 4pt ,line join=round,line cap=round] (293.67, 73.20) -- (311.67, 73.20);
\definecolor{drawColor}{RGB}{0,0,0}

\node[text=drawColor,anchor=base west,inner sep=0pt, outer sep=0pt, scale=  1.00] at (320.67, 81.76) {$Pr(\delta = 0) = 0.5\quad$};

\node[text=drawColor,anchor=base west,inner sep=0pt, outer sep=0pt, scale=  1.00] at (320.67, 69.76) {$Pr(\delta = 0) = 0.1\quad$};

\node[text=drawColor,anchor=base,inner sep=0pt, outer sep=0pt, scale=  1.00] at (114.36,119.50) {$\sigma = 0.8$};

\node[text=drawColor,anchor=base,inner sep=0pt, outer sep=0pt, scale=  1.00] at (114.36,239.32) {$\sigma = 0.24$};

\node[text=drawColor,anchor=base,inner sep=0pt, outer sep=0pt, scale=  1.00] at (121.29,173.15) {$\sigma$ uncertain};
\end{scope}
\end{tikzpicture}
\caption{Comparing the scaled assurance for fixed variance parameters (upper and lower pairs of curves), and with the elicited prior distribution for the variances (middle pair of curves), for the two scenarios corresponding to different probabilities of no treatment effect. In this example, the scaled assurance is sensitive to different fixed values of the variance, but less sensitive to the mass placed on the event of no treatment effect.}\label{scaledAssurance}
\end{figure}

\section{Multi-stage trials}

Given the elicited distributions, we can then investigate what information a proposed trial would provide about $\delta$. Here, we consider a scenario in which a small trial will be conducted, and then a decision will be made whether to commit to a larger trial.  The trial planner would want to know how informative the small trial would be; whether it resolve uncertainty about $\delta$ sufficiently to make the decision about the larger trial easier.

We suppose the trial sponsor chooses some threshold $c$ of interest. From the expert's elicited distribution we will have $Pr(\delta > c) = x$, which is prior to a proposed small trial. We now consider whether the small trial would resolve this uncertainty: given the data $D$ that the trial would produce, whether $Pr(\delta > c|D)$ would be close to either 0 or 1. Before the study is conducted, we do not yet know what the data $D$ would be, and so we think of
$Pr(\delta > 0|D)$ as a random variable: a function of the unknown data $D$.

The expert's prior distribution $\pi(\delta,\sigma^2_c, \sigma^2_t)$ will imply a predictive distribution for $D$, given specified numbers of patients $n_t$ and $n_c$ in the treatement and control arms. Without loss of generality, we can assume $\mu_c=0$ so that $\mu_t = \delta$. We can use the following simulation algorithm to explore the distribution of $Pr(\delta > 0|D)$. 

\noindent\textbf{Algorithm 2: simulating the information gained from a trial}\\
Inputs: sample sizes $n_t$ and $n_c$, the elicited prior $\pi(\delta, \sigma^2_t, \sigma^2_c)$, and the number of iterations $N$. \\
For $i=1,\ldots,N$:
\begin{enumerate}
\item sample $\delta_i,\sigma^2_{t, i}$ and $\sigma^2_{c,i}$ from $\pi(\delta, \sigma^2_t, \sigma^2_c)$;
\item sample $x_{1,i},\ldots,x_{n_t,i}$ from $N(\delta_i, \sigma^2_{t,i})$ and $y_{1,i},\ldots,y_{n_c,i}$ from $N(0, \sigma^2_{c,i})$;
\item define $D_i = (x_{1,i},\ldots,x_{n_t,i},y_{1,i},\ldots,y_{n_c,i})$;
\item using Markov chain Monte Carlo, generate a sample $\delta_{i,1},\ldots,\delta_{i,M}$ from the posterior distribution of $p(\delta|D_i)$;
\item estimate $Pr(\delta > c |D_i)$ by
$$
\hat{P}r_i = \frac{1}{M}\sum_{j=1}^M I(\delta_{i,j}>c),
$$
\end{enumerate}
where $I()$ is the indicator function. This produces an (approximate) sample $\hat{P}r_1,\ldots,\hat{P}r_N$ from the distribution of $Pr(\delta > c|D)$. We can then inspect the sample to see how many probabilities are close to either 0 or 1. We use \texttt{rjags} \citep{rjags2018} to implement the MCMC sampling, and will comment on the choice of $N$ and $M$ in the following example.

\subsection{Example}

We now give an illustration, continuing the example from Section \ref{sec:Example}.  We consider the prior given by $Pr(\delta = 0) = 0.5$, with conditional distribution $\delta | \delta \neq 0 \sim N(0.4, 0.22^2)$ and $\sigma_t^{-2},\sigma_c^{-2}\stackrel{i.i.d}{\sim}Gamma(shape = 2.27,\, rate = 0.29)$. Hence, prior to the small study, we have $Pr(\delta >0) = 0.48$.

We consider a study with $n$ patients per arm, and wish to assess how much more confident we would be that $\delta>0$, given the study data $D$. For illustration, we will classify a study as `informative' if, once the study has produced data $D$, we would have either  $Pr(\delta>0|D)>0.95$ or $Pr(\delta>0|D)>0.95$.

We implement Algorithm 2 with $N=500$ simulated studies, and $M=1000$ generated values of $\delta$ from each Markov chain. The total computation time (for four different values of $n$) was approximately 5 minutes on a desktop computer, using a single core (parallel computation could have been used here). This gives estimates of the probability of an ``informative'' study that are accurate to the first decimal place, which in this context, and noting the reliance on elicited judgements, is likely to be sufficient.

In Table \ref{tableInformative1}, we illustrate the information that could be presented to a decision-maker, to enable a quick comparison of different choices of $n$. A more detailed summary, for $n=20$ is presented in Figure \ref{figureInterim}

\begin{table}[H]
\begin{center}
\begin{tabular}{r|c |c | c | c}
Number of patients per arm & 5 & 10 & 20 & 40 \\
\hline
Probability of `informative' study& 0.2& 0.5 & 0.7 & 0.9
\end{tabular}
\end{center}
\caption{The probability that, following a study producing data $D$ with the specified number of patients per arm, we would either have $Pr(\delta = 0 |D)>0.95$ or  $Pr(\delta = 0 |D)<0.05$}\label{tableInformative1}
\end{table}

\begin{figure}[H]
\begin{center}
\input{interim.tex}
\caption{Distribution of $Pr(\delta = 0 |D)$, as a function of the unknown data $D$ resulting from a trial with 20 patients in each arm. The numbers at the top give the probabilities of $Pr(\delta = 0 |D)$ lying in the respective bins. The distribution suggests that, following such a trial, it is likely we would have little uncertainty as to whether $\delta>0$ or not.}\label{figureInterim}
\end{center}

\end{figure}

This analysis could also be used to provide feedback about the elicited priors: in some cases certain results may be judged implausible, suggesting a problem with the choice of prior. Specifically, we can investigate the probability that a study with one or two patients per arm would be ``informative": one might judge that such a probability should be close to 0. For illustration, we compare two priors:
\begin{itemize}
\item Prior 1: $Pr(\delta = 0) = 0.5$, with conditional distribution $\delta | \delta \neq 0 \sim N(0.4, 0.22^2)$ and $\sigma_t^{-2},\sigma_c^{-2}\stackrel{i.i.d}{\sim}Gamma(shape = 2.27,\, rate = 0.29)$.
\item Prior 2: $Pr(\delta = 0) = 0.5$, with conditional distribution $\delta | \delta \neq 0 \sim N(0.4, 0.22^2)$ and $\sigma_t^{-2},\sigma_c^{-2}\stackrel{i.i.d}{\sim}Gamma(shape = 43.86,\, rate = 0.82)$. 
\end{itemize}
The first prior is the same as that used in the previous example. The gamma prior for $\sigma_t^{-2}$ in Prior 2 results from using the same method in Section \ref{sec:method}, but now supposing that the proportion of patients would would not benefit from the treatment would be between 0.05 and 0.1, implying smaller values of $\sigma_t^2$. 

Table \ref{tableInformative2} shows the estimated probabilities of an `informative' study, for each prior and different (small) values of $n$. Under the belief that $\sigma_t^2$ will be small, it would only take one observed response in the treatment group moderately above $\mu_c$ to `persuade' us that $\delta >0$. Assuming this result is implausible, we would then revisit the elicited priors.

\begin{table}[H]
\begin{center}
\begin{tabular}{r|c |c | c }
Number of patients per arm & 1 & 2 & 5 \\
\hline
Prior 1: probability of `informative' study& 0.00 & 0.01 & 0.02\\
Prior 2: probability of `informative' study& 0.41 & 0.50 & 0.70

\end{tabular}
\end{center}
\caption{The probability that, following a study producing data $D$ with the specified number of patients per arm, we would either have $Pr(\delta = 0 |D)>0.95$ or  $Pr(\delta = 0 |D)<0.05$ \label{tableInformative2}. Arguably, the probabilities are implausibly high for Prior 2, suggesting a problem with the elicited prior distribution.}
\end{table}

\section{Summary\label{sec:conclusion}}
We have proposed a method for eliciting a distribution about a variance parameter, and have illustrated how to incorporate this within an assurance calculation in clinical trial planning. Making judgements about variability within a population is likely to be difficult, but our method does at least avoid asking an expert to update her beliefs given hypothetical data, or to provide summaries from her predictive distribution which would require `mentally integrating out' uncertain parameters. 

If a fixed value of a variance parameter were to be used in an assurance calculation, the assurance could be highly sensitive to the choice of fixed value, so incorporating uncertainty about variances should result in assurance calculations that are more robust. Accounting for variance uncertainty will also be important if there is interest in assessing how much information a small study is likely to produce.

\appendix

\section*{Appendix}

An R package, \texttt{assurance}, for implementing the methods described in this paper is available on GitHub, at \texttt{https://github.com/OakleyJ/assurance}. The website also includes an illustration of using the package to replicate the examples in this paper.

This package currently requires the development version of the \texttt{SHELF} R package, also available on GitHub. These packages can be installed with the commands.

\begin{knitrout}
\definecolor{shadecolor}{rgb}{0.969, 0.969, 0.969}\color{fgcolor}\begin{kframe}
\begin{alltt}
\hlkwd{install.packages}\hlstd{(}\hlstr{"devtools"}\hlstd{)}
\hlstd{devtools}\hlopt{::}\hlkwd{install_github}\hlstd{(}\hlstr{"OakleyJ/SHELF"}\hlstd{)}
\hlstd{devtools}\hlopt{::}\hlkwd{install_github}\hlstd{(}\hlstr{"OakleyJ/assurance"}\hlstd{)}
\end{alltt}
\end{kframe}
\end{knitrout}

An app for implementing these methods \citep[produced with \texttt{shiny},][]{Chang:2019} can be used online at \texttt{https://jeremy-oakley.shinyapps.io/assurance-normal/}. A version of the app for offline use is included in the \texttt{assurance} package.

\bibliography{refdatabase}
\bibliographystyle{rmd}

\end{document}